# An overview about Networks-on-Chip with multicast support

*Marcelo Daniel Berejuck*
Software and Hardware Integration Laboratory
Federal University of Santa Catarina
Rua Pedro João Pereira, 150, 88.905-120 - Araranguá SC - Brazil
marcelo.berejuck@ufsc.br

***Abstract*—Modern System-on-Chip (SoC) platforms typically consist of multiple processors and a communication interconnect between them. Network-on-Chip (NoC) arises as a solution to interconnect these systems, which provides a scalable, reusable, and an efficient interconnect. For these SoC platforms, multicast communication is significantly used for parallel applications. Cache coherency in distributed shared-memory, clock synchronization, replication, or barrier synchronization are examples of these requests. This paper presents an overview of research on NoC with support for multicast communication and delineates the major issues addressed so far by the scientific community in this investigation area.**

## I. Introduction

The silicon industry has used Systems-on-Chip with multiple heterogeneous processing units as means to deliver the performance required by modern applications. However, the integration of an increasing number of specialized processing units poses a challenge on the interconnection mechanisms in such systems. As a solution, the silicon industry has been using Networks-on-Chip to interconnect components in this kind of SoC [1], [2].

The Communication in NoC can be either unicast or multicast [3]. In unicast communication, a message is sent from a processor connected to the network to a single destination processor. In multicast communication, a message is sent from one processor to an arbitrary set of destination processors in the network.

Several SoCs with multiple processors have applications that employ multicast communication, for example, barrier synchronization, cache coherency in distributed shared-memory architectures, or clock synchronization. Despite the multicast communication can be implemented by multiple unicast communications, it is not a suitable strategy because it degrades the performance and increases the congestion in the network [4].

This paper presents an overview of research on NoC with support for multicast communication and delineates the major issues addressed by the scientific community in this research area over the last decade. It starts with an overview about Network-on-Chip and Multicast in Section II. Section III introduces an overview of research that has done on multicast communication for Networks-on-Chip, and Section IV close this paper with our conclusion.

## II. Background

### A. *Network-on-Chip basics*

A NoC consists of a structure of routers and links implementing a packet-switched communication fabric, and Figure 1 depicts an example of 2x2 NoC with four nodes. A node is defined here as a set composed of a router, a network adapter and one core as highlighted in Figure 2(c). Network adapter (NA) acts as a bridge from the packet-switched interconnect to a core connected to the network. For MPSoC based on NoC, a core is typically a processor with some amount of local memory. In Figure 1, IM and DM represent instruction and data memories, which can be a cache or a scratch-pad memory. The core can also be a slave device, such as a controller for a larger amount of external memory shared by several cores, or an Input / Output controller.

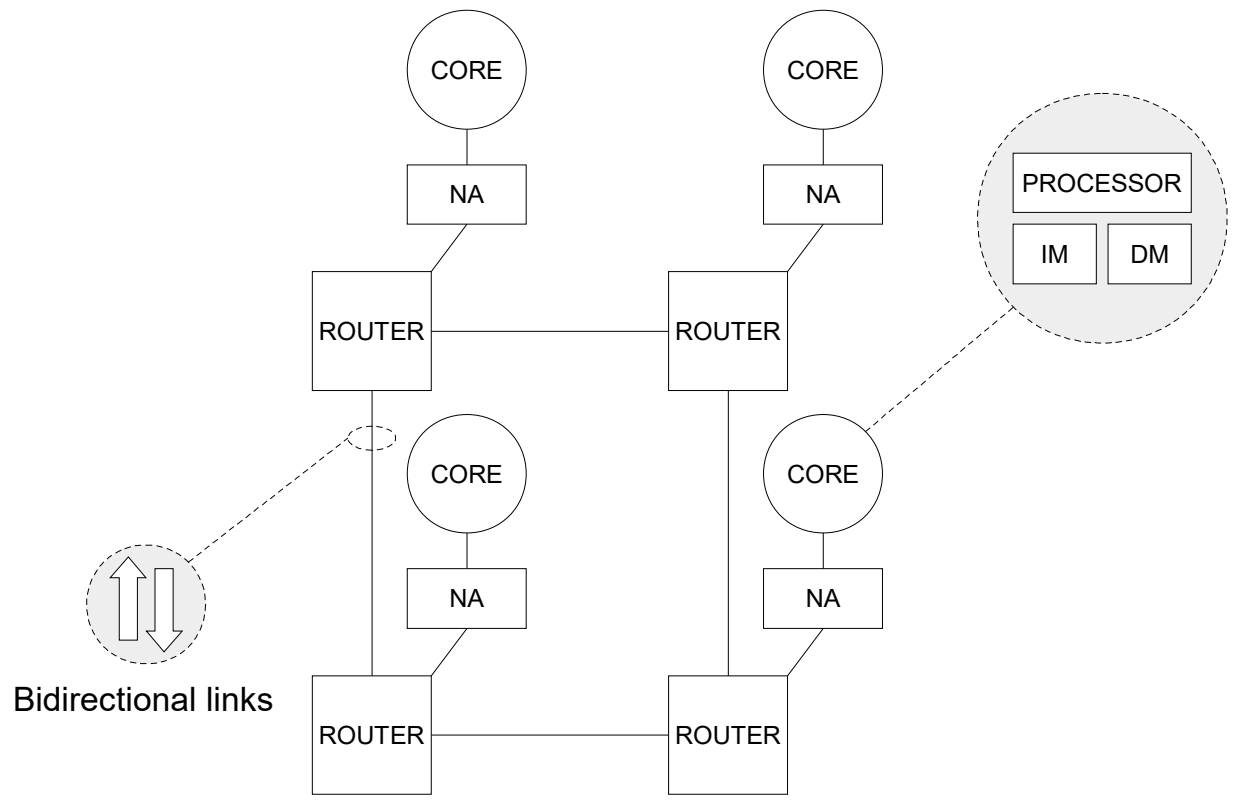

Figure 1. Example of a NoC with four nodes.

Processing cores attached to the NoC communicate by sending and receiving messages through the network. A header, a payload, and a trailer compose messages. The header flit has routing information that guides the packet through the network up to the destination node. The payload to routers along the path that it is the last flit of the packet, and other packets can use the resources (i.e. router and links).

A Network-on-Chip can be described by its topology and the strategies employed for routing, switching, flow control, arbitration and buffering. A network topology is the arrangement of nodes and channels into a graph, as depicted in Figure 2. A Network may have a regular topology, as shown in Figure 2(a), (b), and (c), or irregular topology illustrated in figure 2(d).

Router, highlighted as a white square on the top of Figure 2(c), implements the routing, switching, flow control, arbitration, and buffering functionalities. Routing determines how a message chooses a path in the network. Switching is the mechanism that removes data from an input channel and places it on an output channel. Flow control deals with the allocation of channels and buffers to a message as it traverses this path. Arbitration is responsible for scheduling the use of channels and buffers by the messages. Finally, buffering defines the approach used to store messages while the router arbitration circuits cannot schedule them.

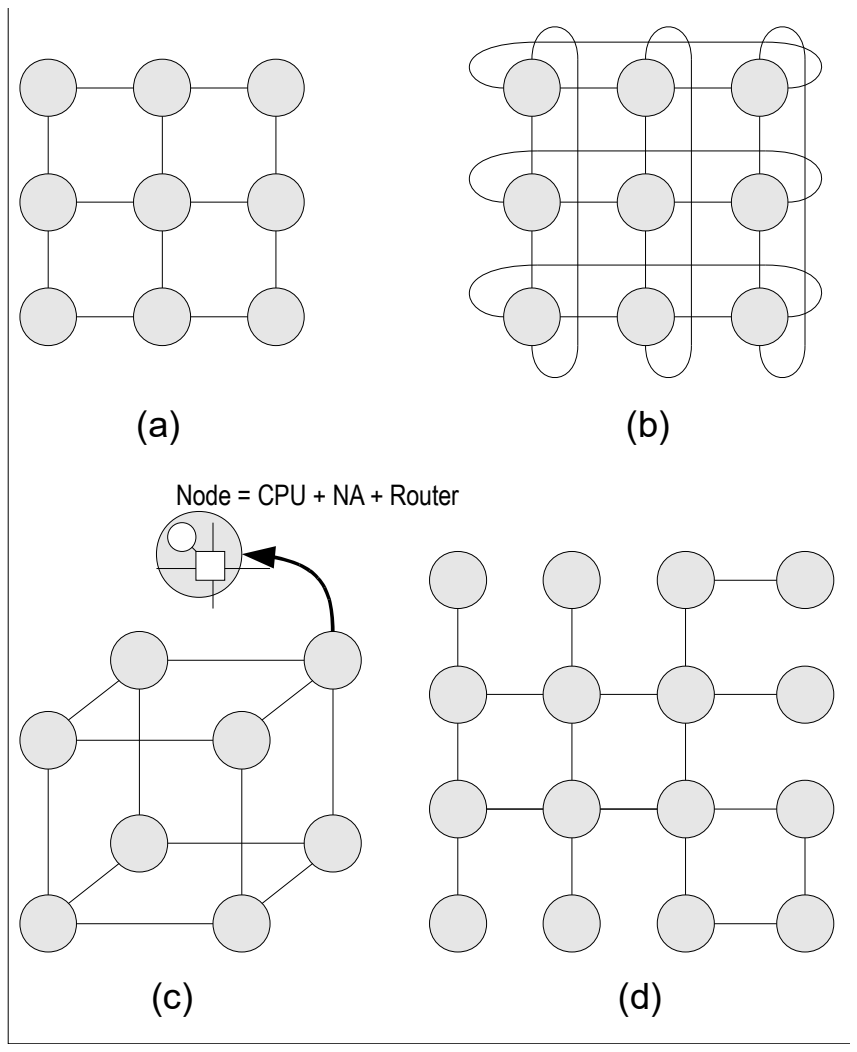

Figure 2. Example of NoC topologies: (a) n-mesh; (b) torus; (c) hypercube; and (d) irregular mesh.

## B. Multicast basics

Extensively research have done on multicast communication issues in computer networks and interconnection networks [11]. Intra-chip interconnections based on Network-on-Chip has a different set of requirements and hence demand different approaches than those employed in computer networks or interconnection networks. Usually, these requirements are related to communication latency, silicon area, and power consumption. Authors as [11] understand that an efficient multicast approach for NoCs should result in low network latency, low power, and low silicon area consumptions.

According to the authors [3], the multicast communication can be classified as unicast-based, path-based, and tree-based. In unicast-based, the multicast operation is performed by sending a separate copy of the message from the source to every destination or, alternatively, by sending the unicast message to a subset of destinations [7]. As a side effect, this scheme suffers from a significant network latency and high power consumption [4].

In the path-based multicast, a message is forwarded to each destination sequentially and following one path [3]. This multicast scheme is attractive due to its simplified hardware design. As a trade-off, path-based multicast may increase the network latency when compared with unicast-based or tree-based multicast [7]. From switching technology point of view, the multicast path-based communication is more suitable for wormhole switching [11].

In tree-based multicast approach, the tree is "virtually" constructed in which the source can be seen as the root and then messages are sent down the tree [7]. The idea is to deliver a message along with a common path as far as possible. Messages can be replicated, as branches, for a single set of destination nodes when needed. If one of those branches gets blocked, all other also will be blocked. In this case, the branches must proceed forward in lock step, which may cause a message to hold many channels for extended periods, resulting in increasing network contention [7]. The tree-based multicast approach is suitable for networks that employ store-and-forward or virtual cut-through routing. For wormhole networks, the tree-based approach incurs high congestion [4].

Our purpose in this Section was to give a highlight about Network-on-Chip concepts, further introduce an overview of multicast issues. Next Section introduces the researches on multicast for NoC presented in the last decade.

# III. Multicast solutions proposed for NoC

## A. Deadlock-Free Multicast Routing Algorithm for Wormhole-Switched Mesh NoC

Authors [5] introduced a deadlock-free adaptation of the dual-path multicast algorithm originally proposed by [8] for multicomputer, targeting wormhole-switched 2-D mesh topology. The algorithm adapted by [5] was implemented on a circuit switching Network-on-Chip.

According those authors, the reason to choose a circuit switching was to send only one copy of the message, independently from the number of destinations. They proposed a routing algorithm based on Hamiltonian paths [6] used for both unicast and multicast messages. It was done to eliminate deadlock in the network since there was compatibility on the messages routing.

The deadlock avoidance was achieved replicating the consumption channels. That feature enabled the injection of several simultaneous multicast messages into the network.

## B. Dynamic multicast routing protocol for distributing traffic in NOCs

Authors [7] proposed a method to take advantage of the network partitioning, and utilizing of a destination ordering algorithm. In the proposed scheme, three points of view have been considered. First is the utilization of network partitioning, and second is the optimized destination ordering. Third is taking advantage of the odd-even turn model adaptive algorithm for routing both the multicast and unicast messages through the network.

They employed an adaptive routing algorithm that used the congestion condition of the input ports to route messages through non-congested paths, to distribute the load to preventing highly congested area problem.

According to those authors, the proposed algorithm has good behavioral under high message injection rates. They state that the algorithm had the lowest average communication delay in comparison with the Dual-Path [8], Multi-Path [8], and Column-Path [9] multicast routing algorithms.

### *C. Tree-based multicast routing for write invalidation messages in networks-on-chip*

Authors [10] proposed a multicast router for a single-flit write invalidation message in a Network-on-Chip. In their approach, the destination sets were partitioned by the four directions of the router and all destination nodes are reached through the shortest path.

A single packet is injected into the network to cover all destinations, and the packet is replicated to be forwarded to requested outputs at a router in the packet path. The write invalidation message proposed by them consists of a single-flit packet using a bit string encoding for the multidestination routing header.

If the destination node is out of range of a reachable network, a level of indirection is used where the destination node is covered with a unicast packet which has a binary destination tag. It means that the proposed router supports two types of destination address encoding schemes: binary encoding of destination identification for unicast routing and bit-string for multicast routing.

### *D. Recursive partitioning multicast*

Authors [11] proposed a recursive partitioning multicast routing and a multicast wormhole router design for Networks-on-Chip. According to those authors, the RPM routing allows routers to select intermediate replication nodes based on the global distribution of destination nodes, providing more path diversities and achieving more bandwidth efficiency.

Router latency affects the packet delivery latency. As a workaround, those authors used a router design technique based on look-ahead routing and speculative switch allocation, to reduce the number of pipeline stages. Look-ahead routing removes the routing computation stage from the pipeline by making a routing decision one hop ahead of the current router.

Speculative switch allocation enables the virtual channel allocation stage to be performed with the switch allocation stage simultaneously. A separate switch allocator finds available input and output ports of the crossbar after the normal switch allocator reserves them. Consequently, the router proposed by [13] have only two stages.

### *E. Customized Networks-on-Chip Architectures With Multicast Routing*

Authors [12] proposed a formulation of the custom Network-on-Chip synthesis problem based on the decomposition of the problem into the inter-related steps of finding a good set partition of traffic flows. That strategy generates a good physical network topology for each group in the partition and providing an optimized network implementation for the derived topologies.

The problem formulation had taken into consideration the unicast and multicast traffic. The authors proposed four algorithms for systematically examining different possible set partitioning of flows. Their framework design flow integrates floor planning, and they described several ways to ensure deadlock-free routing of both unicast and multicast flows.

### *F. NoC Multi-casting Scheme in Support of Multiple Applications Running on Irregular Subnetworks*

Authors [13] introduced an irregular sub-network oriented multicast routing strategy. Such strategy was based on the following assumption: if the output channel found by regular topology oriented multicast routing is not available, therein the routing strategy could choose an alternative output channel. This strategy also leads to the minimal path to the destination.

Those authors suggested that following this strategy: an irregular topology oriented multicast routing algorithm could be designed based on any regular mesh-based multicast routing algorithm. They proposed an algorithm called Alternative Recursive Partitioning Multi-casting, based on Recursive Partitioning Multi-casting algorithm, which was designed for regular mesh topology originally.

The basic idea of their algorithm was to support multicasting for irregular sub-networks based on Recursive Partitioning Multicasting, which only supports multi-casting for regular mesh topology. The target was to find the output directions following the basic Recursive Partitioning Multi-casting algorithm and then decide to replicate the packets to the original output directions or the alternative output directions based on the shape of the network.

### *G. Multicast messages in cache-coherence protocols for NoC-based MPSoCs*

Authors [14] proposed the use of multicast messages to reduce the number of transactions to improve the performance of cache coherence protocols in Network-on-Chip based Multiprocessor System-on-Chip (MPSoC). They use a homogeneous MPSoC platform called HeMPS [15] to evaluate their hypothesis. It was described in synthesizable VHDL, in which processing elements are connected to a Network-on-Chip called Hermes [16]. An MIPS-like processor called PLASMA, a local memory RAM, a DMA controller and a Network Interface composes each processing element.

The proposed work adopted two-level memory hierarchy. The first level is the shared memory accessible by all processing elements. It contains a memory controller responsible for executing read/write operations according to the cache-coherence protocol, and a directory memory that stores for each block its status and the address of processing elements holding a copy of it.

The results introduced by those authors suggest the effectiveness to employ low-level NoC services, as multicast, to optimize the cache coherence protocol.

### *H. A TDM NoC with support to multicast communication*

Authors [17] proposed a circuit switching network which supports multicast and offers hard guarantees regarding bandwidth and latency per connection. The network uses a time-division multiplexing (TDM), contention-free scheme and a distributed routing model similar to one of the Æthereal [18].

They implemented the network configuration mechanism as a dedicated broadcast network with a tree topology, with links running in parallel to a subset of the normal data network links. One core called "host" has the control over the configuration infrastructure through a configuration module connected to a node where the host is connected.

The subset of links that build the configuration tree is chosen in such a way as to minimize the distance from the host to any of the network nodes. The configuration infrastructure is employed to set up data connections by updating the contents of the TDM slot tables inside routers and network interfaces. It configures and reads back the state of the network interfaces.

### *I. Planar Adaptive Router Micro-architecture for Tree-Based Multicast Network-on-Chip*

Authors [19] proposed a scheduling mechanism for tree-based multicast routing using deadlock free static and partially planar adaptive routing algorithms. They were addressing the multicast deadlock configuration problem in the intermediate nodes of a NoC they called XHINoC.

In proposed routing, multicast packets are routed and scheduled in the NoC using a local identity-based multiplexing technique with wormhole switching. They used an identity tag attached to every flit, which allows different flits of different packets to be mixed in the same queue and enables to implement a fair flit-by-flit round arbitration to share communication links.

According to those authors, their methodology can guarantee lossless flit acceptance in multiple destination nodes even for long size multicast messages. Furthermore, they ensure that there is also no out-of-order delivery problem, due to the dedicated working organization of the combined router hardware logic and routing look-up table units.

### *J. NoC router supporting multicast*

Authors [20] proposed a deadlock-free routing algorithm focus on the deadlock issues on Networks-on-Chip when it involves multicast communication. They assumed the topology of the network as 2D mesh with m columns and n rows.

Each node of the network has a label L, with its assignment function, expressed regarding nodes coordinates. The routing algorithm uses a differentiated sub-network strategy. The labels virtually divide the routers of the network into two sub-networks, called "up sub-network" and "down sub-network." Each router consists of two sub-routers, called "up sub-router" and "down sub-router." All the up sub-routers compose the up sub-network, and this network a sub-router can only forward packets that have a label bigger than it does.

As the network resources are divided into two subsets and in each sub-network multicast packets share fewer network resources in their transmission directions, those authors understand that the possibility of deadlock is decreased. Furthermore, when a deadlock is formed, the related multicast packets will be aware of and turn to the only transmission path in horizontal directions, thus avoiding the deadlock condition.

### *K. Summary of the related work*

This Section introduced the multicast solutions proposed for network-on-Chip in the last decade. This subject has been addressed since 2008. After six years, we found few papers focusing on multicast in NoC and all of them introduced in this Section. Table I summarizes the related work with support for multicast on Network-on-Chip that we analyzed.

TABLE I. SUMMARY OF NETWORK-ON-CHIP WITH MULTICAST SUPPORT.

| Author | Year | Multicast | Problem addressed | Strategy adopted |
|---|---|---|---|---|
| [5] | 2008 | Path-based | Deadlock in Network-on-Chip with several simultaneous multicast messages. | Adaptation of the dual-path multicast algorithm originally proposed by [8] for multicomputer and circuit switching. |
| [7] | 2009 | Path-based | Deadlock in Network-on-Chip with several simultaneous multicast messages. | Take advantage of the network partitioning, and utilizing of a destination-ordering algorithm. |
| [10] | 2009 | Tree-based | Network latency and power consumption on a multicast Network-on-Chip. | A multicast router for a single-flit write invalidation message in the network. |
| [11] | 2009 | Path-based | Network latency and power consumption on a multicast Network-on-Chip. | A recursive partitioning multicast routing algorithm and a multicast wormhole router design. |
| [12] | 2009 | Tree-based | Network contention and silicon consumption. | Formulation of the custom Network-on-Chip synthesis problem based on the decomposition of the problem into the inter-related steps of finding a good set partition of traffic |
| | | | | flows. |
| [13] | 2011 | Tree-based | Multicast on Network-on-Chip with irregular topology. | An irregular topology oriented multicast routing algorithm based on a regular mesh-based multicast routing algorithm. |
| [14] | 2011 | Path-based | Performance of the memory organization on an MPSoC based on NoC. | Use of multicast messages to reduce the number of transactions in order to improve the performance of cache coherence protocols in Network-on-Chip based Multiprocessor System-on-Chip. |
| [18] | 2012 | Tree-based | Silicon cost and compromising the guarantees of service on a NoC with multicast support. | A network configuration mechanism as a dedicated broadcast network with a tree topology, with links running in parallel to a subset of the normal data network links. |
| [19] | 2012 | Tree-based | Multicast deadlock configuration problem in the intermediate nodes of a NoC. | Extra bits added to every flit used as a local identity-based multiplexing technique with wormhole switching. |
| [20] | 2014 | Tree-based | Complex deadlock in NoC when it involves multicast communication. | Routing algorithm that check node labels, such that virtually it divides the routers of the network into two sub-networks. |

# IV. Conclusion

The proposal of this paper was to present an overview of research on NoC with support for multicast communication and delineated the major issues addressed so far by the scientific community in this research area. Although it presents just an overview of all NoC research related to NoCs with multicast communication, it reveals the main issues and how the researchers deal with that. We noticed that 80% of the research presented so far have focused on routers structure of the networks in terms of algorithms [5] [7][10][11][13][20], adding parallel links working in the network [18], or extra bits into the flits structure [19]. Other 20% of research have focused on the design-time network partition [12], and message scheme was running on the processing cores connected in the network [14]. We also noticed that few papers are focusing on multicast on NoC have published so far.

We understand that it is a quite new research area, and the interest on it may increase due to the increasing on multi-core Systems-on-chip. The reason for that increase is the continuous advances in the semiconductor device fabrication technology that expected to reach fourteen nanometres for silicon fabric [21]. The continued downscaling of the silicon technology increases transistor density and operating frequency, and hence the SoC demands lower supply voltage, which reduces noise immunity. At this point, Network-on-Chip emerges as a suitable interconnection mechanism; it is recognized as a solution for electrical issues on SoCs composed by several processing cores [22], what may expand the research horizon for multicast communication on NoCs.

The Future scope of this research work will be a detailed survey, in which, the technical details of each research introduced in this paper will be mitigated. It will serve as a base for an evaluation that our team intends to do with different multicasting techniques on the RTSNoC[23], a Network-on-Chip designed for real-time communication flows.